\documentstyle[11pt]{article}

\makeatletter
\long\def\@makecaption#1#2{
 \vskip 10pt
 \setbox\@tempboxa\hbox{{\small\bf#1:} \small#2}
 \ifdim \wd\@tempboxa >\hsize {\small\bf#1:} \small#2\par
 \else \hbox to\hsize{\hfil\box\@tempboxa\hfil}
 \fi}
\makeatother

\def\medfigsize{9 truecm}
\def\medcapsize{4.8 truecm}

\def\twofigsize{6.9 truecm}
\def\twocapsize{6 truecm}

\font\smalltitlefont = cmssdc10 scaled \magstep 3
\font\sectionfont    = cmssdc10 scaled \magstep 2
 1

\makeatletter

\def\maketitle{
	\par
	\begingroup
		\def\thefootnote{\fnsymbol{footnote}}
		\def\@makefnmark{\hbox to 0pt{$^{\@thefnmark}$\hss}}
		\if@twocolumn \twocolumn[\@maketitle]
		\else
			\newpage
			\global\@topnum\z@ \@maketitle
		\fi
		\thispagestyle{plain}
		\@thanks
	\endgroup
	\setcounter{footnote}{0}
	\let\maketitle\relax
	\let\@maketitle\relax
	\gdef\@thanks{}
	\gdef\@author{}
	\gdef\@title{}
	\let\thanks\relax
	}

\def\@maketitle{
	\newpage
	\null
	\vskip 2em
	\begin{center}
		{\smalltitlefont \baselineskip 20pt \@title \par}
		\vskip 1.5em
		{
			\large \lineskip .5em
			\begin{tabular}[t]{c} \@author \end{tabular}
			\par
		}
		\vskip 1em
		{\large \@date}
	\end{center}
	\par
	\vskip 1.5em
	}

\def\section{\@startsection {section}{1}{\z@}{-3.5ex plus -1ex minus
 -.2ex}{2.3ex plus .2ex}{\sectionfont}}

\makeatother

\newcommand{\const}{{\rm const.}}

\makeatletter

\def\vereq#1#2{\lower3pt\vbox{\baselineskip1.5pt \lineskip1.5pt
	\ialign{$\m@th#1\hfill##\hfil$\crcr#2\crcr\sim\crcr}}}

\makeatother

\def\antiLambda{{\overline\Lambda}}

\input epsf   

\def\today{\number\day \
     \ifcase\month\or Jan\or Feb\or Mar\or Apr\or May\or June
	 \or July\or Aug\or Sep\or Oct\or Nov\or Dec \fi \
     \number\year}

\begin{document}

\title{A hydrodynamical assessment of 200 A GeV collisions}

\author{Ekkard Schnedermann$^a$ and Ulrich Heinz$^b$
   \\ [6 pt]
   $^{(a)}$Physics Department, Brookhaven National Laboratory\\[-2pt]
   Upton, New York 11973, USA \\
   and \\
   $^{(b)}$Institut f{\"u}r Theoretische Physik,
   Universit{\"a}t Regensburg\\[-2pt]
   D-93040 Regensburg, Germany \\
   }
\date{\today}

\maketitle

\begin{abstract}
\noindent
We are analyzing the hydrodynamics of $200\,A\,\rm GeV$ S+S collisions
using a new approach which tries to quantify the uncertainties arising
from the specific implementation of the hydrodynamical model. Based on
a previous phenomenological analysis we use the global hydrodynamics
model to show that the amount of initial flow, or initial energy
density, cannot be determined from the hadronic momentum spectra. We
additionally find that almost always a sizeable transverse flow
develops, which causes the system to freeze out, thereby limiting the
flow velocity in itself. This freeze-out dominance in turn makes a
distinction between a plasma and a hadron resonance gas equation of
state very difficult, whereas a pure pion gas can easily be ruled out
from present data. To complete the picture we also analyze particle
multiplicity data, which suggest that chemical equilibrium is not
reached with respect to the strange particles. However, the
overpopulation of pions seems to be at most moderate, with a pion
chemical potential far away from the Bose divergence.
\end{abstract}

\section{Introduction}
\label{introduction}

The success of hydrodynamics as a model for the space-time evolution
of heavy-ion collisions at the BEVALAC with center-of-mass energies of
several hundred MeV/nucleon \cite{Stoecker,BEVALAC} suggests its basic
applicability to the ultrarelativistic heavy-ion collisions at the
Brookhaven AGS with $\sqrt{s}\approx 5\,A\,\rm GeV$ or at the CERN SPS
with $\sqrt{s}\approx 20\,A\,\rm GeV$ (e.g.
\cite{Katajahydro,Katajaflow,Gersdorffhydro,Ornikhydro,Katscherhydro}).
Since the hydrodynamical model
\cite{Fermi,Landau,Carruthershydro,Bjorken} describes the matter
during the collision in terms of the collective variables energy
density $\varepsilon$, pressure $P$ and flow velocity $u^\mu$, its
validity should actually improve at higher energies because of the
strongly increasing particle multiplicities. Although the model's
primary task is the description of the inclusive hadron spectra, which
are measured in several experiments at the AGS
\cite{AGSdata,borlaenge} and the SPS \cite{borlaenge,SPSdata}, its
reasonably precise picture of the collison dynamics can furtheron
serve as a basic framework for the computation of almost any other
observable, e.g.\ $J/\psi$, strangeness, $\overline p$, dileptons,
strangelets, etc.

Moreover, such an effort is also accompanied by the hope that once a
reliable hydrodynamical description is achieved we will automatically
have information about the yet unknown equation of state of hadronic
matter, which serves as the basic input to hydrodynamics.  Especially,
a first order phase transition from normal hadronic matter to a quark
gluon plasma, as suggested by lattice QCD calculations, would result
in a coexistence region with $\partial P/\partial \varepsilon \to 0$
and, in principle, might lead to qualitatively new dynamical
signatures.

However, we encounter the problem that some of the assumptions
underlying the hydrodynamic model  may not be sufficiently accurate in
ultrarelativistic collisions with  center-of-mass energies one order
of magnitude higher than at the BEVALAC. In particular, the question
whether and when local thermalization can be reached is still under
debate, and we do not attempt to answer it in this paper.  Instead we
will take sufficient thermalization for granted some time after the
first hard collisions and only assume validity of the hydrodynamic
model thereafter. Apart from thermalization, our approach will differ
considerably from previous publications
\cite{Katajahydro,Katajaflow,Gersdorffhydro,Ornikhydro,Katscherhydro}
in that we will try to resolve the other important uncertainties
arising from the specific implementation of hydrodynamics. To achieve
this goal we are careful to test each hypothesis when it is
introduced, leading to the organization of the paper as follows.

Based on a phenomenological analysis of the hadronic spectra from
$S$+$S$ at $200\,A\,\rm GeV$ \cite{pheno}, we extract in
Sec.~\ref{thedatafromahydrodynamicalpointofview} from the measured
particle spectra as much information as possible about the
hydrodynamical behaviour at freeze-out. In
Sec.~\ref{extractingtheinitialconditions} we use global hydrodynamics
as a computationally efficient method \cite{global}, to investigate to
what extent the initial hydrodynamical state is constrained by the
data and find that the amount of initial longitudinal flow cannot be
decided from the data alone. A closer look at two possible extreme
scenarios in Sec.~\ref{fullstoppingvspartialtransparency}, namely
Landau- and Bjorken-like initial conditions, shows striking
similarities in their temperature and transverse flow at freeze-out.
In Sec.~\ref{longitudinalandtransversefreezeouttimescales} we show
that the transverse expansion is the most important contributor to the
freeze-out process and is thus invariably limited by itself.  In
Sec.~\ref{comparisonswiththedata} we confront our model with the
experimental data and note good agreement for a variety of underlying
assumptions regarding entropy generation and freeze-out description.
Additionally, we test different equations of state and find that only
the pure pion gas can be ruled out while the distinction between the
plasma and the hadron resonance gas by hydrodynamical means is very
hard to accomplish.  In Sec.~\ref{particleratiosmultiplicities} we
finally address the issue of chemical equilibration and the pion
chemical potential by computing particle ratios and multiplicities.
Sec.~\ref{summaryandconclusions} summarizes our results and
conclusions.

\section{The data from a hydrodynamical point of view}
\label{thedatafromahydrodynamicalpointofview}

The measured single particle distributions confer basically two
distinct types of information: The shape of the measured spectra
reflects the temperature and flowdynamics of the system, while the
absolute normalization of the spectra is connected additionally with
the size of the collision zone and the amount of chemical
equilibration among the particle species.

\begin{figure}[htb]
	\begin{minipage}[t]{\twofigsize}
		\epsfxsize \twofigsize \epsfbox{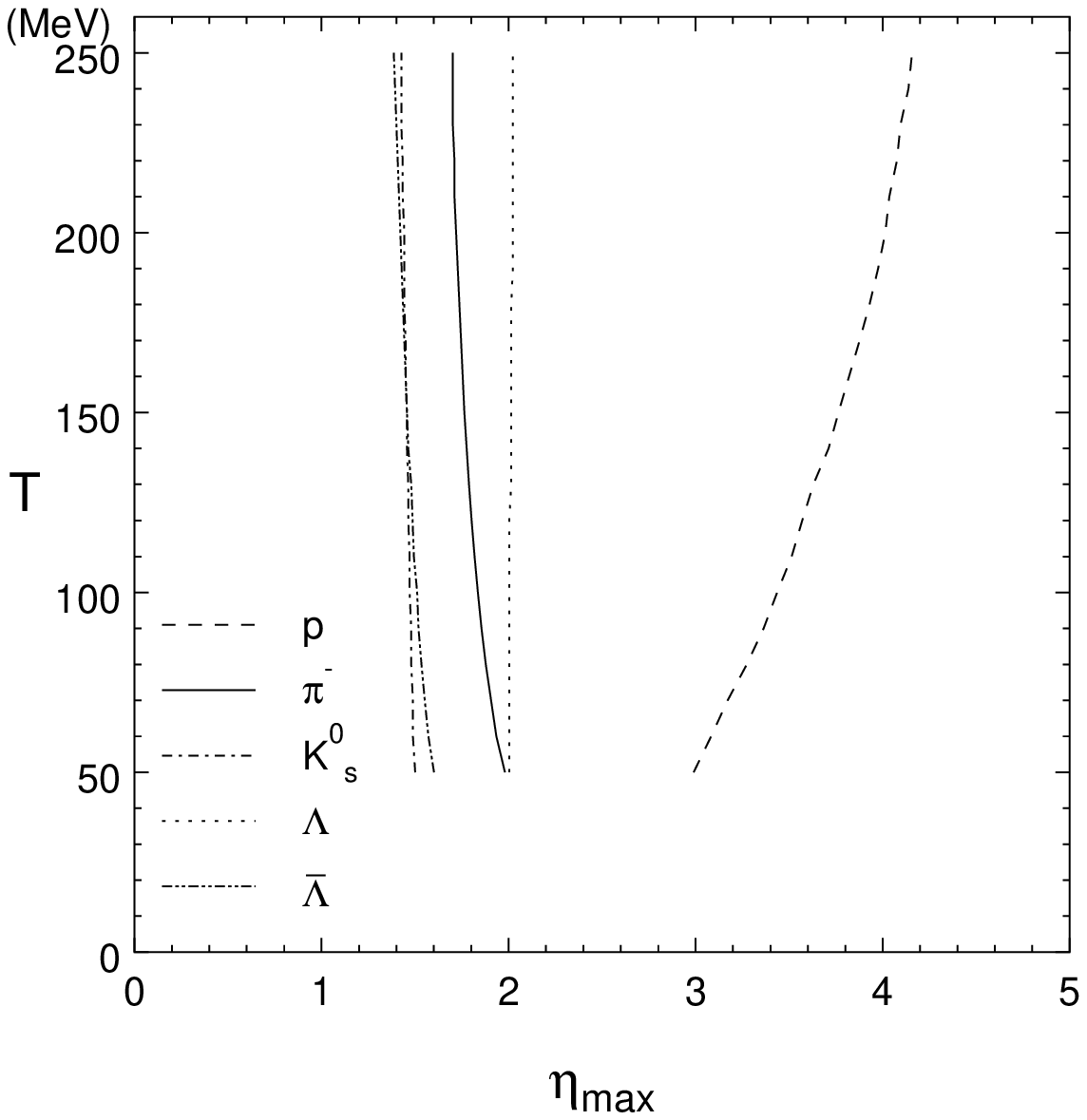}
		\hspace*{0.5truecm}
		\begin{minipage}{\twocapsize}
		\caption[]{ \label{fityall} \sloppy
			    Every point $(T,\eta_{\rm max})$ on these lines corresponds to a
			    fit of the computed $y$-spectra (including resonance decays and
			    longitudinal flow) to the measured ones
    \cite{NA35strange,WenigD,NA35charged}. The longitudinal flow
    $\eta_{\rm max} \approx 1.7\pm 0.2$ fits all particle spectra
			    almost independently of the temperature $T$, with exception of the
			    protons, which still carry a big amount of their initial
    longitudinal motion~\cite{pheno}.
			}
		\end{minipage}
	\end{minipage}
	\hfill
	\begin{minipage}[t]{\twofigsize}
		\epsfxsize \twofigsize \epsfbox{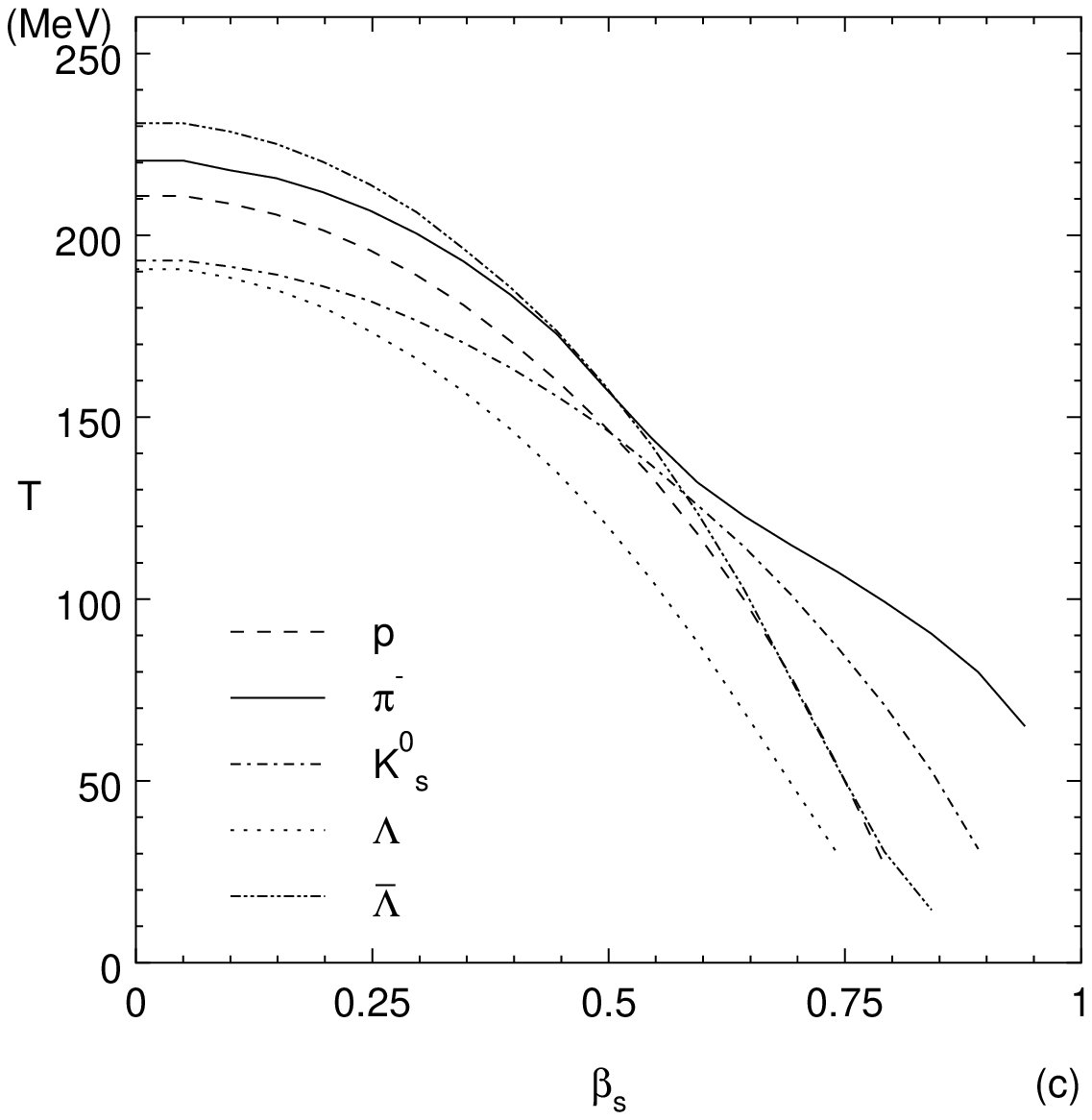}
		\hspace*{0.5truecm}
		\begin{minipage}{\twocapsize}
		\caption[]{ \label{fit2all} \sloppy
			    Every point $(T,\beta_s)$ on these lines corresponds to a fit of
			    the computed $m_T$-spectra (including resonance decays and
			    transverse flow) to the measured ones
    \cite{NA35strange,WenigD,NA35charged}. Temperature $T$ and
			    transverse flow $\beta_s$ have similar effects on the spectra so
    that 			$\beta_s$ can not be extracted unambiguously~\cite{pheno}.
			}
		\end{minipage}
	\end{minipage}
\end{figure}

For the moment we will focus exclusively on the shape, leaving the
normalization arbitrary until Sec.~\ref{particleratiosmultiplicities}
where we return to this issue in detail. Since at SPS energies pions
are the overwhelming majority of all produced secondaries, with kaons,
nucleons, strange baryons and antibaryons following in descending
order, we will base our analysis preferentially on the pions. This
ensures on the experimental side that the spectra can be measured
quite accurately with small statistical errors while on the
theoretical side the understanding of different effects on the spectra
has been developed the furthest. In any case, because the other
particles seem to follow the pions (Fig.~\ref{fit2all}), all species
would give very similar results.

As shown in \cite{pheno}, the dynamics of the collision zone shows up
in the hadronic spectra, as measured by the NA35 collaboration
\cite{NA35strange,WenigD,NA35charged}, in two distinct ways: The
longitudinal flow determines the rapidity distributions, while the
transverse flow influences the transverse mass spectra.

The longitudinal flow can be extracted most accurately from
the pion rapidity spectrum~\cite{Katajaflow,Sinyukovflow,pheno}.  The
extracted value is supported by the spectra of all other particles
measured by NA35~\cite{NA35strange,WenigD,NA35charged}. The rapidity
spectra of all produced particles are well described by a boost
invariant distribution around the center of the collision zone at
$y_{\rm cm}=3$, with the flow rapidities limited to the interval $(-
\eta_{\rm max},+\eta_{\rm max})$ with $\eta_{\rm max} = 1.7\pm
0.2$~\cite{pheno} (see Fig.~\ref{fityall}).

The situation is less clear for the transverse mass spectra, which are
affected both by resonance decays \cite{resonances} and transverse
flow. We found that for any given radial fluid velocity $0\le
\beta_s\le 0.7 \,c$ it is always possible to find a temperature
$T(\beta_s)$, which fits the spectra for all particle species
simultaneously~\cite{pheno} (see Fig.~\ref{fit2all}). The reason is
that with transverse flow the spectra can in good approximation be
described in terms of an {\em apparent} or blueshifted temperature
\cite{pheno}
 \begin{equation}
    T_{app} = T \sqrt{ {1+\langle \beta_r \rangle } \over
                       {1-\langle \beta_r \rangle }  }\ ,
 \label{Tapparent}
 \end{equation}
from which the temperature $T$ and average radial flow $\langle
\beta_r\rangle$ cannot be easily separated.

The amount of transverse flow thus can not be determined from the
particle spectra alone. We have shown \cite{circumstantial} that the
method of global hydrodynamics \cite{global} combined with a realistic
freeze-out criterium is well suited to give ``circumstantial
evidence'' for a transverse flow of $\beta_s \approx 0.5\,c$
\cite{averageflow}. Here we will present a more thorough theoretical
analysis to gain a better understanding of the dynamics of the
collision zone. The method of global hydrodynamics is ideal for such
an analysis, because on the one side it contains all necessary
elements for a consistent description of the dynamics (coupling of
longitudinal and transverse flow by the hydrodynamic equations as well
as a dynamical freeze-out criterium) and on the other side, due to its
numerical effectiveness, it allows for an in-depth phenomenological
analysis including extensive parameter studies.

\section{Extracting the Initial Conditions}
\label{extractingtheinitialconditions}

Hydrodynamics connects the thermalized initial state with the frozen-
out final state. With its help the freeze-out conditions
(Tab.~\ref{initialtab}), reflected in the observed particle spectra,
can be related to the initial conditions and provide new insights into
the dynamical evolution. Some of the initial conditions are known
from the size of the projectile nucleus and the measured energy loss
of the projectile protons \cite{Stroebele,WenigD}. At the freeze-out
point we know the maximal fluid rapidity (Fig.~\ref{fityall}) as
extracted from the $y$-spectra. The $m_\perp$-spectra only yield a
correlation between the transverse fluid velocity $\beta_s$ and the
local temperature $T$ at freeze-out (Fig.~\ref{fit2all}), which is
rather difficult to exploit and the discussion of which we postpone
until Sec.~\ref{comparisonswiththedata}. In principle, it is also
possible to determine the final radius using pion interferometry
(HBT)~\cite{HBT}. Unfortunately, up to now the theoretical and
experimental uncertainties in the ana\-lysis of the 2-pion (and
2-kaon) correlation functions are still too large ($> 0.5$ fm) to
render them very useful for our purpose~\cite{UliMayer}.

\begin{table}[htb]
\begin{center}
\begin{tabular} {|c||c|c||c|c|} \hline
\phantom{$\bigg\vert_N$}
quantity     & initial state  & from:      & final state   & \ \ \ from: \ \ \
\\ \hline\hline
\phantom{$\bigg\vert_N$}
$R$          & $4 \,{\rm fm}$ & projectile size & ?        & (HBT) \\ \hline
\phantom{$\bigg\vert_N$}
$\beta_s$    & 0              & defined    & $(\beta_s,T)$ &
$\frac{dn}{m_Tdm_T}$ \\ \hline
\phantom{$\bigg\vert_N$}
$Z$          & from $E_{\rm tot}$ & $E_{\rm loss}=313\,\rm GeV$ & ?   & ?  \\
\hline
\phantom{$\bigg\vert_N$}
$\eta$       & ?              & ?          & 1.7           & $\frac{dn}{dy}$
\\ \hline
\phantom{$\bigg\vert_N$}
$\varepsilon$& ?              & ?          & ?             & ?     \\ \hline
\end{tabular}
\begin{minipage}{12 truecm}
\caption[]{\sloppy \label{initialtab}
      The initial and final state parameters of the global
      hydrodynamics are only partially known. Through hydrodynamics
      we can combine our knowledge about the initial and final state
      and constrain the unknown parameters.
      }
\end{minipage}
\end{center}
\end{table}

\begin{figure}[htb]
	\begin{minipage}[t]{\twofigsize}
		\epsfxsize \twofigsize \epsfbox{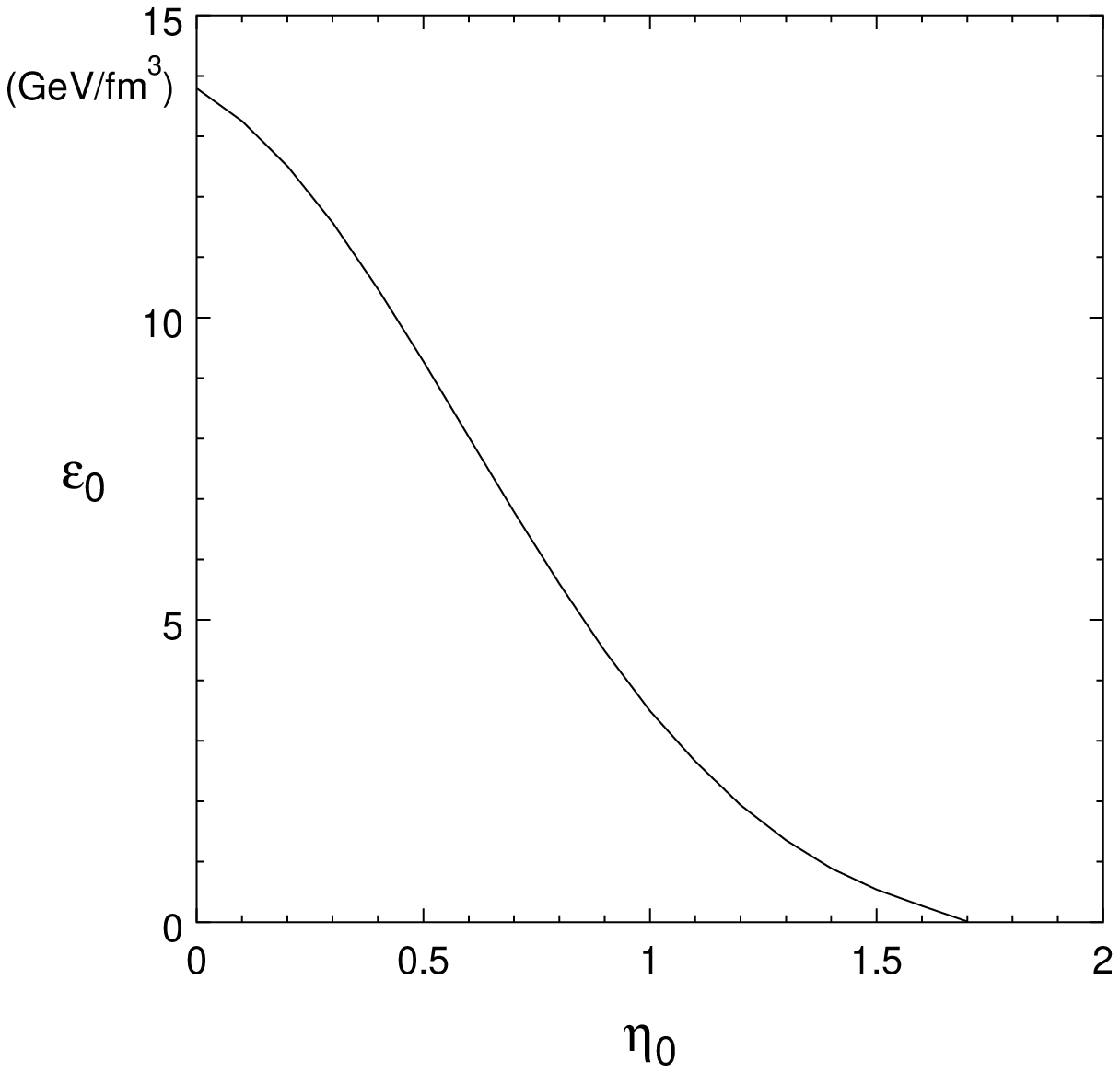}
		\hspace*{0.5truecm}
		\begin{minipage}{\twocapsize}
		\caption[]{\label{eeta} \sloppy
			    Shown is the initial energy density $\varepsilon_0(\eta_0)$ which,
			    for a given initial longitudinal flow $\eta_0$, generates at
			    freeze-out the measured flow $\eta_f = 1.7$. The hydrodynamics is
			    based on a hadron resonance gas EOS with $S/A=30$.
			}
		\end{minipage}
	\end{minipage}
	\hfill
	\begin{minipage}[t]{\twofigsize}
		\epsfxsize \twofigsize \epsfbox{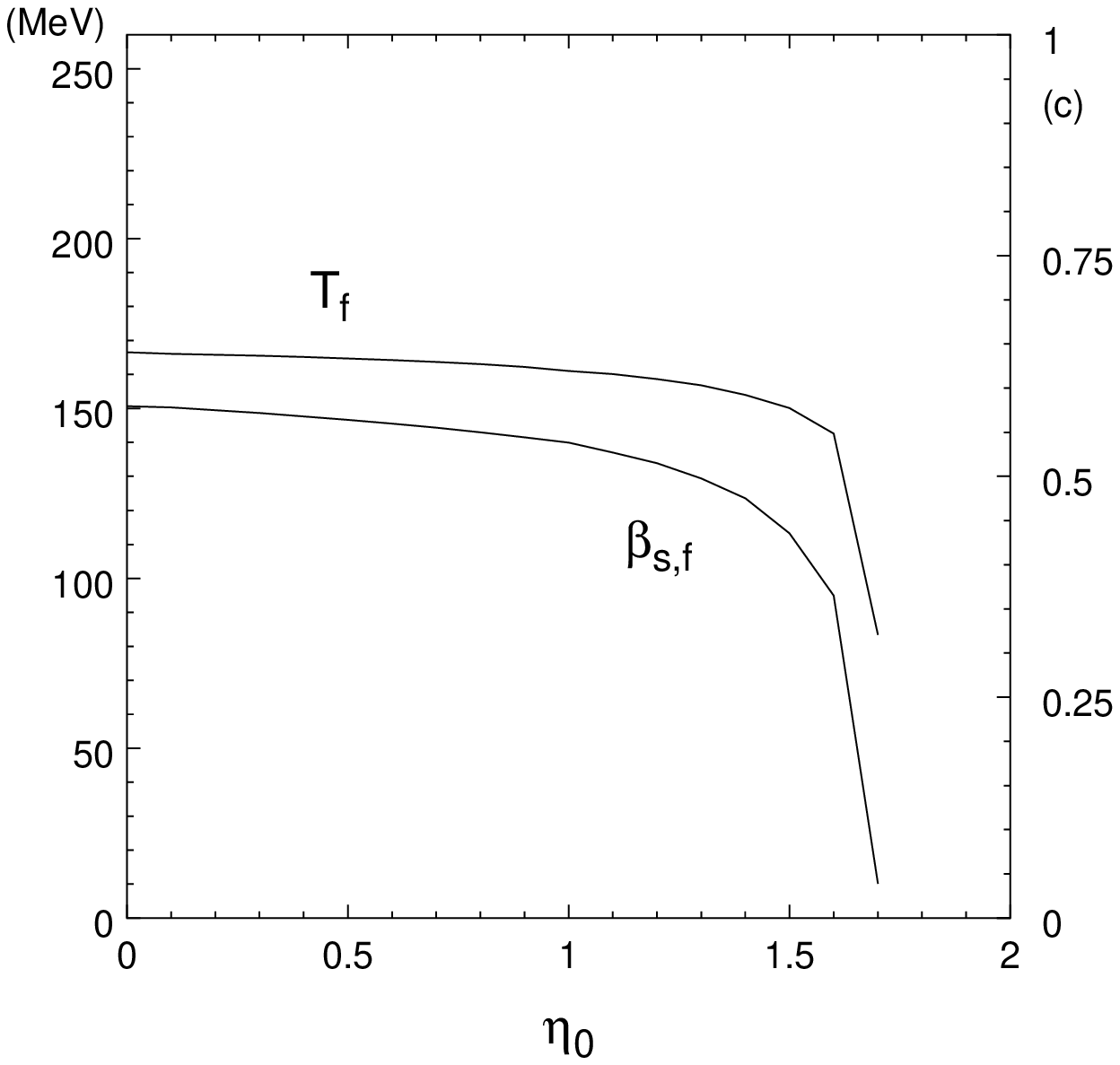}
		\hspace*{0.5truecm}
		\begin{minipage}{\twocapsize}
		\caption[]{\label{Tbetaseta} \sloppy
			    Temperature $T$ and transverse flow $\beta_{s,f}$ at freeze-out as
			    determined from the longitudinal flow $\eta_f=1.7$ via the global
			    hydrodynamic model. Both are almost constant over a whole 			range
    of initial conditions for the longitudinal flow, $\eta_0\le 1.6$.
			}
		\end{minipage}
	\end{minipage}
\end{figure}

The idea of this paper is to use the global hydrodynamics framework
of Ref.~\cite{global} to fill in (most of) the question marks in
Tab.~\ref{initialtab} by establishing a connection between the initial
and the final state. We proceed strategically and first narrow down
the uncertainties in the initial conditions by choosing from all
combinations of initial longitudinal flow $\eta_0$ and initial energy
density $\varepsilon_0$ only those pairs $(\eta_0,\varepsilon_0)$
which after hydrodynamical evolution produce at freeze-out a
longitudinal flow value of $\eta_f = 1.7$.  In Fig.~\ref{eeta} we show
the result for a hadron resonance gas equation of state with specific
entropy $S/A=30$.

For the case of complete stopping (no longitudinal flow at the
beginning, $\eta_0=0$) we need very high energy densities
$\varepsilon_0 > 10 \,\rm GeV/fm^3$, so that the strong longitudinal
pressure gradient can accelerate the matter to $\eta_f=1.7$ at
freeze-out. For larger initial flows considerably smaller initial
energy densities suffice to reach the same freeze-out value. In the
extreme case of very large initial flows $\eta_0 \approx \eta_f$ the
system is already close to freeze-out initially, i.e. the initial
energy densities are very low, and the hydrodynamics barely generates
additional longitudinal flow.

Knowing the possible pairs of initial conditions $(\varepsilon_0,
\eta_0)$, we can now study the systematics of the hydrodynamically
generated transverse flow at freeze-out, parametrized by $\beta_s$,
within the allowed range of longitudinal flow parameters.
Surprisingly, for a large region of initial conditions, ranging from
complete stopping to considerable initial transparency ($0\le \eta_0
\le 1.6$), more or less the same (sizeable) amount of transverse flow
$\beta_s\approx 0.5\,c$ is generated \cite{averageflow}. Above $\eta_0
\ge 1.6$ no real hydrodynamical evolution takes place since the matter
is already close to freeze-out and is accelerated very little.
Although strictly speaking we cannot exclude this possibility within
the framework presented here, we will disregard this region from now
on, because there the assumption of local equilibrium which underlies
the hydrodynamic model breaks down, and a non-equilibrium description
\cite{nonequilib} appears to be more suitable.

The temperature at freeze-out turns out to be similarly insensitive to
the initial longitudinal flow.  This is largely a consequence of the
constancy of $\beta_{s,f}$, since the temperature at freeze-out is
coupled to the transverse velocity (see
Sec.~\ref{longitudinalandtransversefreezeouttimescales}) via the
scattering time scale $\tau_{\rm sca}(T)$ and the expansion time scale
$\tau_{\rm exp}(R, Z, \eta_f, \beta_{s,f})$ such that higher
velocities cause the system to freeze out already at higher
temperatures and densities.

{}From this we can already conclude that with hadronic observables alone
the initial energy density and the initial longitudinal flow can not
be determined. All measurable quantities at freeze-out are
essentially independent of $\eta_0$ (and thus of $\varepsilon_0$).
Other obervables, for example lepton pairs~\cite{HydroLepton}, might
prove useful to determine the initial energy density: due to their
small cross sections they freeze out immediately and thus the measured
spectrum receives contributions from the whole space-time region of
the collision zone, with the hot early stages dominating the spectrum
due to the $T^4$-dependence of the Stefan-Boltzmann law for thermal
radiation.

\section{Full stopping vs. partial transparency}
\label{fullstoppingvspartialtransparency}

We will postpone the application of these results for the moment in
order to analyse in this section in more detail their dynamical
origin.  We pick two extreme cases of the initial longitudinal flow
which we call, according to their resemblance to the Landau and the
Bjorken hydrodynamics, the L-scenario ($\eta_0=0$) and the B-scenario
($\eta_0=1.3$), respectively. Their initial and freeze-out values are
given in Tab.~\ref{initialLB}.

\begin{table}[htb]
\begin{center}
\begin{tabular} {|lr||c|c||c|c|} \cline{3-6}
\multicolumn{2}{c|}{}           &\multicolumn{2}{c||}{L-scenario}
&\multicolumn{2}{c|}{B-scenario} \\ \cline{3-6}
\multicolumn{2}{c|}{}           &initial& final         &initial&final   \\
\cline{3-6}\hline
$R$             &($\rm fm$)     &$4$    &$4.71$         &$4$    &$4.57$	 \\
\hline
$\beta_s$       &($c$)          &$0$    &$0.58$         &$0$    &$0.50$  \\
\hline
$Z$             &($\rm fm$)     &$0.23$ &$4.0$          &$2.3$  &$7.2$   \\
\hline
$\eta$          &               &$0$    &$1.7$          &$1.3$  &$1.7$   \\
\hline
$\varepsilon$   &$\!\!\!\!\!\!\!\!\!\!\rm (GeV/fm^3)$
				&$13.8$ &$0.36$         &$1.4$  &$0.26$  \\ \hline
$T$             &($\rm MeV$)    &$313$  &$167$          &$209$  &$157$   \\
\hline
$S/A$           &               &$30$   &$30$           &$30$   &$30$    \\
\hline
\end{tabular}
\begin{minipage}[t]{11truecm}
\caption[]{\sloppy \label{initialLB}
		    Initial and freeze-out values of the L and the B-scenario, which
		    are distinguished by their initial longitudinal flow $\eta_0$, as
		    computed with our model from the known initial values
		    (Tab.~\ref{initialtab}) and the choice $S/A=30$ for a hadronic
		    EOS.
		}
\end{minipage}
\end{center}
\end{table}

The L-scenario generates the largest transverse flow, as one would have
intuitively expected. But the excess of transverse flow compared to
the B-scenario is rather small; at first sight this is quite
puzzling, considering the large difference in the thermally generated
longitudinal flow $\eta_f-\eta_0$, which is 1.7 for the L-system and
only 0.4 for the B-system.

\begin{figure}[htb]
	\begin{minipage}[b]{\medfigsize}
		\epsfxsize \medfigsize \epsfbox{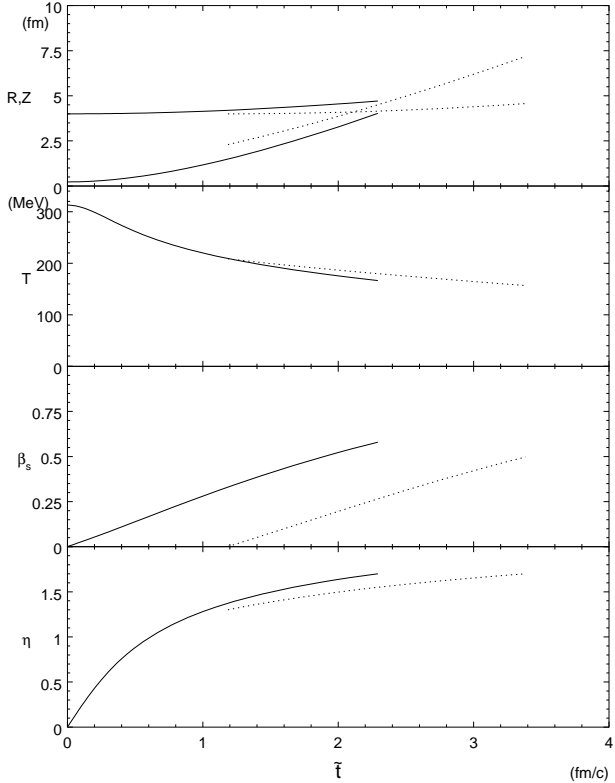}
	\end{minipage}
	\hfill
	\begin{minipage}[b]{\medcapsize}
		\caption[] { \label{dyLB} \sloppy
			    Comparison between the L- and B-scenarios shows remarkable
			    similarities if one shifts the initial time of the B-dynamics so
			    that its temperature agrees with the L-scenario at 			some later
    time. The difference in build-up of transverse flow is			 compensated
    by a later freeze-out in the B-scenario.
			}
		\vspace{1 truecm}
	\end{minipage}
\end{figure}

In Fig.~\ref{dyLB} we compare the dynamical evolution of both
scenarios until freeze-out (where the lines end). The initial time of
the B-scenario was displaced by hand in such a way that at that point
the temperatures in the two scenarios agree. One realizes that the
production of longitudinal flow is strongest at the beginning and then
levels off considerably, even when taking into account the exponential
character of $\eta$.

Neglecting small deviations between both scenarios -- which
incidentally might also be caused by the slightly differing
longitudinally comoving coordinate systems in both cases -- the main
difference clearly lies in $\beta_s$. The L-scenario generates
transverse flow from the beginning, so that there is already a flow
velocity of $\beta_s \approx 0.3\,c$ when the B-system starts. In the
further evolution the L-system freezes out earlier whereas the B-system
nearly catches up in transverse flow during the remaining part of its
expansion history. The reason for the similar transverse flow
velocities at freeze-out thus appears to lie more in the freeze-out
criterium than in the initial conditions.

\section{Longitudinal and Transverse Freeze-Out Time Scales}
\label{longitudinalandtransversefreezeouttimescales}

\begin{figure}[htb]
	\begin{minipage}[t]{\twofigsize}
		\epsfxsize \twofigsize \epsfbox{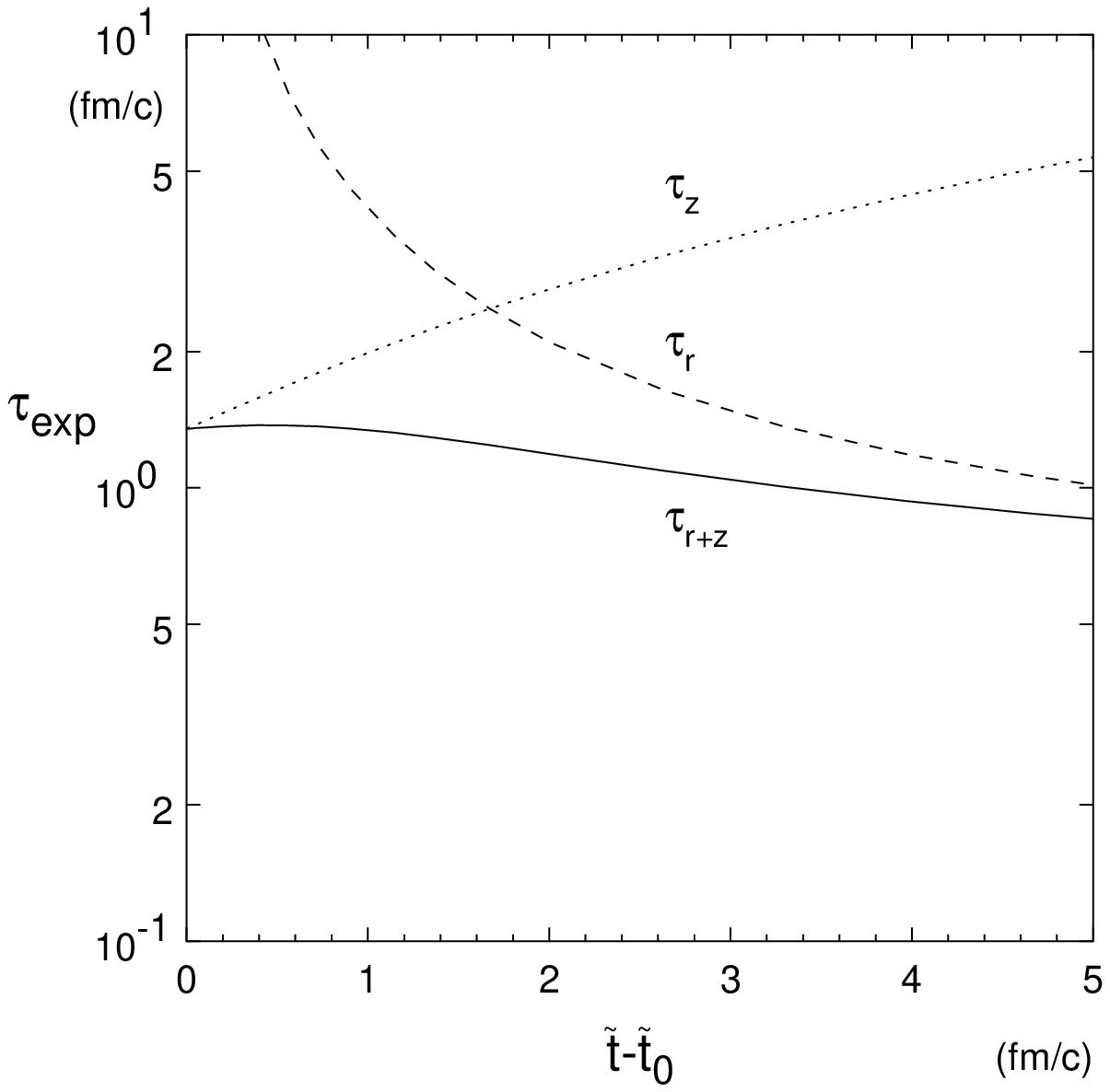}
		\hspace*{0.5truecm}
		\begin{minipage}{\twocapsize}
		\caption[] {\label{tauexp} \sloppy
			    The expansion time scale (here for the B-scenario with
			    $\eta_0=1.3$) is initially dominated by longitudinal expansion.
    After a few ${\rm fm}/c$ the radial expansion takes over. The net
			    result is a rather weak time dependence of the total expansion
    time scale.
}
\end{minipage}
\end{minipage} 	\hfill
	\begin{minipage}[t]{\twofigsize}
\epsfxsize \twofigsize \epsfbox{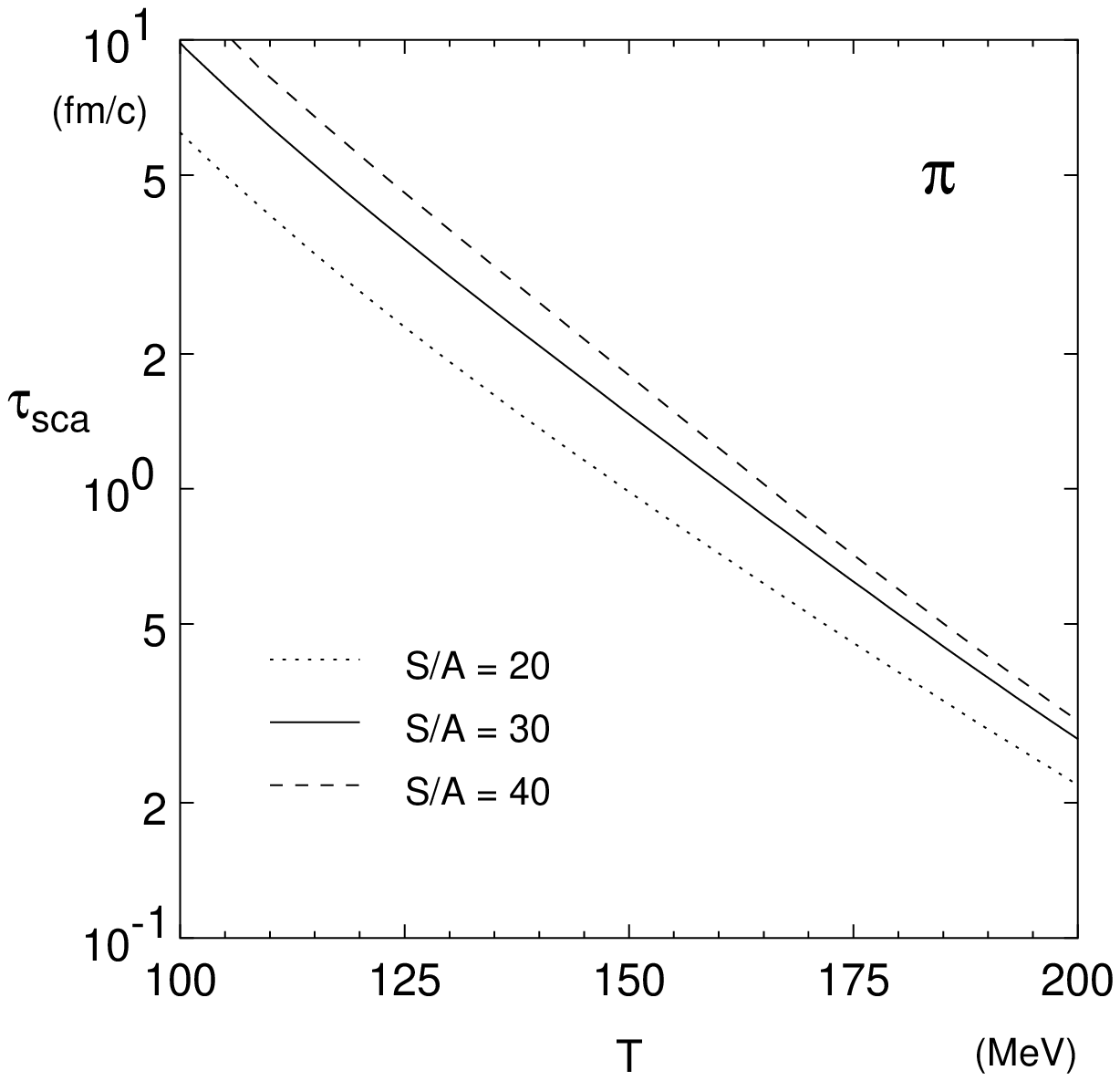}
\hspace*{0.5truecm}
\begin{minipage}{\twocapsize}
		\caption[]{\label{scapi} \sloppy
    The scattering time scale of pions 			in a hadron gas varies with
    $S/A$ because of the 			different baryon content. 			Its exponential
    decrease at higher temperatures, combined with the near constancy
			    			of $\tau_{\rm exp}$, results in 			freeze-out temperatures which are
    always close to $150\rm\,MeV$.
			}
		\end{minipage}
\end{minipage}
\end{figure}

We investigate this question in more detail in Fig.~\ref{tauexp} by
analyzing the dynamically determined expansion time scale $\tau_{\rm
exp}$. Initially the evolution is strongly dominated by longitudinal
expansion, resulting in small values of $\tau_{\rm exp}$. However, the
length of the system quickly increases, reducing the longitudinal
velocity gradients and thus weakening the longitudinal shear forces
which could break the local thermal equilibrium. A simple estimate in
the limit of boost invariant longitudinal expansion gives an expansion
time scale $\tau_{\rm exp} = \tilde t$ which grows linearly with
proper time.

The transverse radius, on the other hand, stays for a long time close
to its initial value. Due to quadratic $R$-dependence of the
transverse area of the cylinder, already rather small transverse
velocities of order $\beta_s\approx 0.4\,c$ cause appreciable dilution
and lead to transverse expansion time scales which can compete with
the longitudinal one. This is illustrated by the non-relativistic
estimate \cite{global} $\tau_{\rm exp} \approx (n+1)\beta_s/R \approx
3\,{\rm fm}/c$, with $n=2$ for a quadratic transverse velocity profile
and $R=4\,{\rm fm}/c$ for a sulphur nucleus. We conclude that the
rarefaction at later stages is strongly dominated by the transverse
flow, which thereby limits its own growth. A comparison with the
scattering time scale $\tau_{\rm sca}$ in Fig.~\ref{scapi}, which
constitutes our freeze-out criterium $\tau_{\rm exp}=\tau_{\rm sca}$
\cite{BondorfGarpmanZimanyi,global}, then shows why always nearly the
same freeze-out temperature $T_f$ is generated.

Our whole analysis is based on this dynamical freeze-out criterium,
which explicitly models the competition between the local expansion,
which disturbs equilibrium, and the scattering processes, which
restore equilibrium. If we had implemented the freeze-out through a
static criterium, e.g. via a freeze-out temperature $T_f=\const$
(typically $140\,\rm MeV$), this kind of analysis would not have been
possible. It is interesting to observe, however, that the result of
dynamical freeze-out nearly coincides with such a criterium, due to
the rapid cooling of the system by transverse expansion and the
exponential temperature dependence of the scattering time scale.

In summary we arrive at the remarkable result that there is no strong
correlation between produced longitudinal and produced transverse
flow, because the former is mainly generated initially while the
latter results from the late dynamical stages before freeze-out. This
also explains retrospectively why, in spite of the big differences
with regard to the longitudinal dynamics, the spherical model of
Ref.~\cite{Search} yields similar values for the transverse flow at
freeze-out.

\section{Comparisons with the Data}
\label{comparisonswiththedata}

We can now confront our theoretical prediction for the relationship
between $\beta_s$ and $T$ at freeze-out with the pairs $(T_f,
\beta_{s,f})$ we have extracted from the phenomenology of the
transverse mass spectra \cite{pheno}. The hope is that this comparison
permits to narrow down the amiguity between thermal and collective
motion from the phenomenology of the transverse momentum data
\cite{pheno}. However, even though the phenomenological analysis of
the data gives an anticorrelation between $T_f$ and $\beta_{s,f}$
according to Eq.~(\ref{Tapparent}), whereas the freeze-out criterium
correlates $T$ and $\beta_{s,f}$ directly (faster expansion causes
freeze-out at higher $T$), it is a priori by no means guaranteed that
the two approaches yield comparable $(T,\beta_s)$ values at all: the
only input for the theoretical hydrodynamical model are the rapidity
distributions, while its results are checked by using the independent
data from the transverse mass distributions.

\begin{figure}[htb]
	\begin{minipage}[b]{\medfigsize}
		\epsfxsize \medfigsize \epsfbox{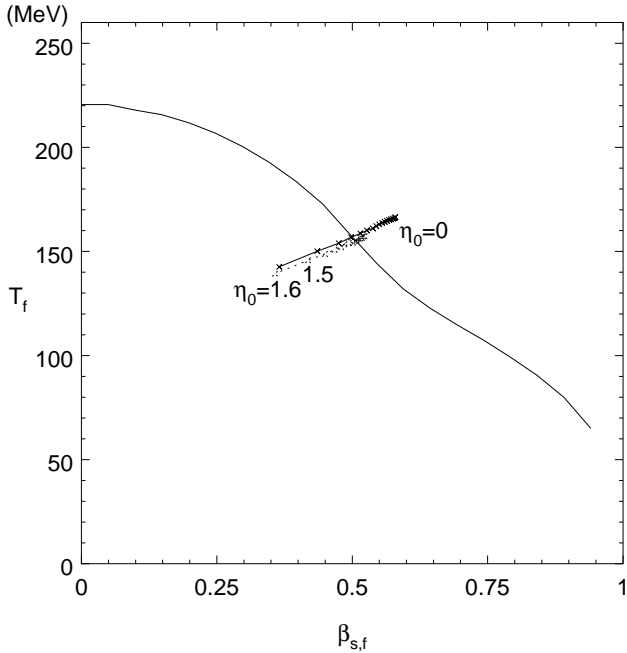}
	\end{minipage}
	\hfill
	\begin{minipage}[b]{\medcapsize}
		\caption[] { \label{Tfbetasf} \sloppy
			    A comparison of the theoretical correlation between temperature
			    and trans\-verse flow at freeze-out with the experimental
			    correlation deduced from the pion spectra (Fig.~\ref{fit2all}).
    The intersection at $\beta_s \approx 0.5\,c$ can be interpreted as
			    evidence for transverse flow in the experiment. The hydrodynamic
			    system with $\approx 10\%$ entropy generation (dotted line,
    \cite{global})			 gives a similar result.
			}
		\vspace{1 truecm}
	\end{minipage}
\end{figure}

For the comparison we combine the theoretically determined
$\beta_{s,f}(\eta_0)$ and $T_f(\eta_0)$ into a line
$(T_f,\beta_{s,f})$, which we parametrize by the unknown initial
longitudinal flow $\eta_0$, covering the range from $0$ to $1.6$ as
indicated next to the line in Fig.~\ref{Tfbetasf}. From
Fig.~\ref{Tbetaseta} we already know that the pairs
$(T_f,\beta_{s,f})$ for all initial conditions with genuine
hydrodynamical evolution, i.e.\ for initial flow rapidities $0\le
\eta_0\le 1.6$, are concentrated in a small region near
$\beta_s\approx 0.5\,c$ and $T\approx 150 \,\rm MeV$.

It is gratifying to see in Fig.~\ref{Tfbetasf} that the
$(T,\beta_s)$-curve extracted from the data crosses this region; this
implies that the phenomenological $(T,\beta_s)$-correlation is
consistent with hydrodynamical evolution and freeze-out, and that our
dynamical picture of the collision region successfully explains the
data. We can use this agreement between theory and data as support for
our assumption of local thermalization and hydrodynamical evolution
and interpret the result as a theory-based proof for the existence of
transverse flow at freeze-out. To assess the credibility of such an
interpretation and to venture into new grounds, we now proceed to
examine our analysis under various additional aspects.

The interest for transverse flow in heavy-ion collisions has been
fueled also by the hope to find an indicator for a phase transition
into the quark gluon plasma \cite{vanHove}. In an intuitive scenario
the initial collision energy in longitudinal direction should be
partially converted into the latent heat associated with a first order
phase transition. The subsequent cooling and decay of the plasma would
then release the energy isotropically to all particles, so that in
particular the transverse momenta would profit.

\begin{figure}[htb]
	\begin{minipage}[b]{\medfigsize}
		\epsfxsize \medfigsize \epsfbox{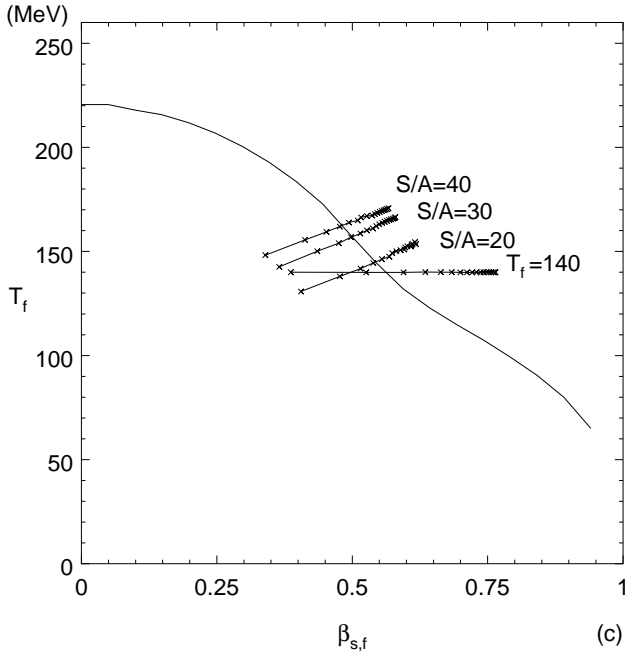}
	\end{minipage}
	\hfill
	\begin{minipage}[b]{\medcapsize}
		\caption[] {\label{Tbetasdiv} \sloppy
			    Sensitivity of the transverse flow to different assumptions:			 The
    reference curve from Fig.~\ref{Tfbetasf} is labelled by $S/A=30$.
			    Other hadron resonance equations of state with $S/A=20$ and $40$
			    result in almost the same transverse flow $\beta_s \approx
			    0.5\,c$, whereas freeze-out at constant temperature $T = 140\,\rm
			    MeV$ leads to a slightly larger transverse flow.
			}
		\vspace{1 truecm}
	\end{minipage}
\end{figure}

\begin{figure}[htb]
	\begin{minipage}[t]{\twofigsize}
		\epsfxsize \twofigsize \epsfbox{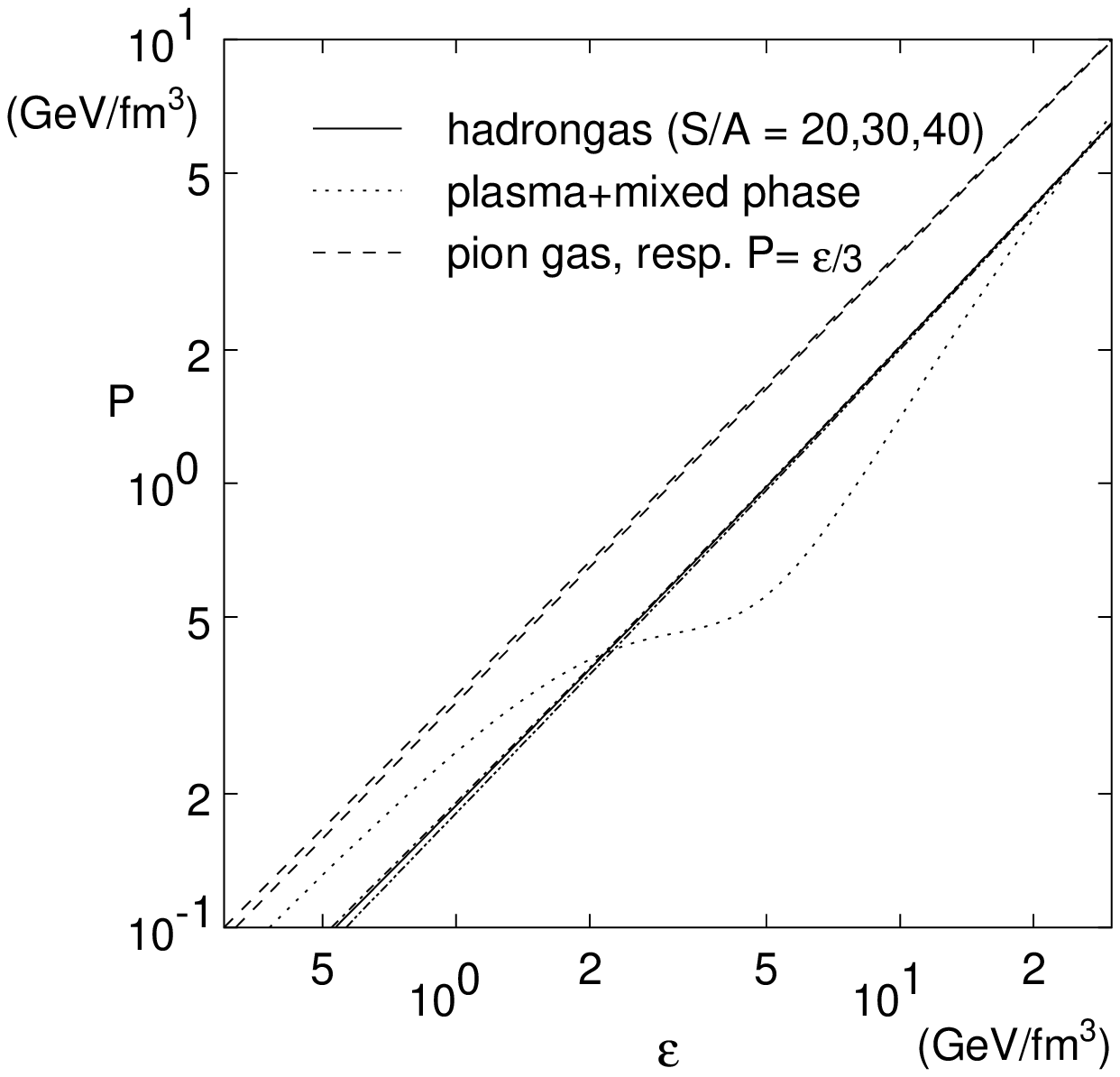}
		\hspace*{0.5truecm}
		\begin{minipage}{\twocapsize}
		\caption[]{\label{eospe} \sloppy
			    The equation of state influences the hydrodynamics by the
			    dependence of the pressure $P(\varepsilon)$ on the energy			 density.
    There are significant differences between a soft hadron resonance
			    gas (solid), the plasma equation of state including a phase
			    transition (dotted) and the hard pure pion gas (dashed).
			}
		\end{minipage}
	\end{minipage}
	\hfill
	\begin{minipage}[t]{\twofigsize}
		\epsfxsize \twofigsize \epsfbox{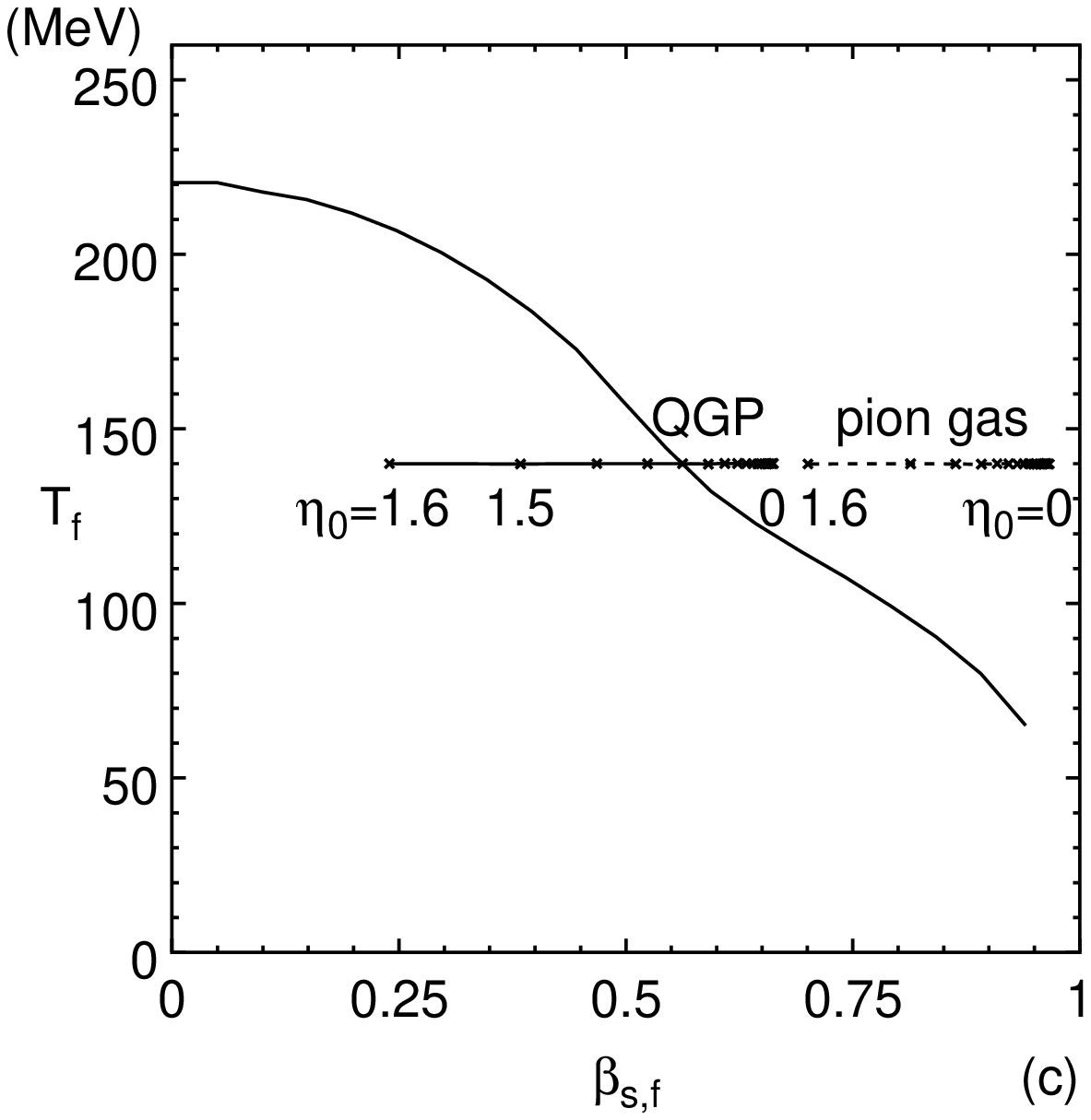}
		\hspace*{0.5truecm}
		\begin{minipage}{\twocapsize}
		\caption[] {\label{Tbetaseos} \sloppy
			    The transverse flow from a plasma EOS differs	 only little from
			    that of a hadron resonance gas when the same 			freeze-out criterium
    is employed (Fig.~\ref{Tbetasdiv}). 			For a pure pion gas EOS all
			    initial conditions with genuine hydrodynamical evolution ($\eta_0\le
    1.5$) generate too much transverse flow and fail to reproduce the
    data.
}
		\end{minipage}
	\end{minipage}
\end{figure}

The possible existence of a phase transition to a quark-gluon plasma
state influences the hydrodynamical evolution through the
corresponding modification of the equation of state of the expanding
matter. We have tested with the global hydrodynamics quantitatively
the equations of state depicted in Fig.~\ref{eospe}. They are
presented in terms of energy density $\varepsilon$ vs. pressure $P$,
the variables entering the hydrodynamical tensor $T^{\mu\nu}$.  We
repeat our procedure to obtain $(T,\beta_s)$ correlations leaving the
unknown initial $\eta_0$ open.

We begin by discussing the hadron resonance gas EOS, which describes
a mixture of relativistic ideal gases of a multitude of hadron
species ($\pi$, $\eta$, $\rho$, $\omega$, $\phi$, $K^+$, $K^0$, $p$,
$n$, $\Delta$, $\Lambda$, $\Sigma$, $\Xi$, $\Omega$, and their
antiparticles). Previously we fixed the entropy per baryon at $S/A
= 30$, to achieve agreement with the measured $p/\pi^-$ ratio
(Sec.~\ref{particleratiosmultiplicities}). We now also include the
choices $S/A=20$ and $40$, corresponding to systems with larger and
smaller baryon densities, respectively. Clearly, all three EOS are
very similar in $P(\varepsilon)$ such that the baryon content has
little influence on the dynamics. However, due to the large
pion-nucleon cross section the baryon content does influence the
scattering time scale (Fig.~\ref{scapi}). The resulting variations
in the freeze-out point can be seen in Fig.~\ref{Tbetasdiv}.

We have also implemented the simple freeze-out criterium $T_f=140$ MeV
= const., as it is often used in hydrodynamics calculations. This also
permits us to compare below the resonance gas and plasma equations of
state, since for the latter the computation of a scattering time scale
for the final state hadrons is not really meaningful except near the
end of the collision. Because the value of 140 MeV is below the
dynamical freeze-out temperatures of our standard scenario, the flow
values in this case tend to be somewhat higher.

An extreme, but popular EOS is that of a gas of non-interacting pions.
It is much harder ($P_\pi(\varepsilon) \gg P_{\rm had}(\varepsilon)$)
than the other equations of state and very close to ideal massless
boson limit $P=\frac{1}{3} \varepsilon$. It thus generates a very
large transverse flow (Fig.~\ref{Tbetaseos}), resulting in a known
(e.g.~\cite{Gersdorffhydro}) disagreement with the observed
$m_T$-spectra. This EOS can thus be excluded by our analysis for all
hydrodynamic initial conditions.

Finally we study a plasma equation of state with a first order phase
transition, where during the mixed phase plasma and hadron gas
coexist. As a representative from this class we have taken the EOS of
U. Ornik \cite{UdoEOS}, which has been obtained by interpolation of
data points from lattice gauge theory calculations. The plasma
EOS exhibits a bend in the coexistence region where the velocity of
sound ${c_s}^2 = \frac{\partial P}{\partial \varepsilon}$ approaches
zero. The comparison of Figs.~\ref{Tbetasdiv} and \ref{Tbetaseos} for
$T_f = 140\,\rm MeV$ shows that the plasma EOS generates a similar
amount of flow as the hadron EOS. This results from the fact that
the transverse flow is generated at later stages, when the system is
no longer in the plasma state but already in the hadron gas phase.

\section{Particle Ratios and Multiplicities}
\label{particleratiosmultiplicities}

While the thermal description of the momentum spectra follows from
the assumption of local thermal equilibrium, which requires only
elastic scattering, a thermal description for the relative particle
multiplicities (e.g. as in \cite{Statistical}) implies also
chemical equilibration which requires additionally inelastic
scatterings. While the former seems to be reasonably valid for
heavy-ion collisions, the latter is much harder to reach in purely
hadronic systems, especially for strange particles \cite{chemequilib}.
In fact, our approach does yield characteristic discrepancies between
theoretical and experimental particle ratios, which indicate that full
chemical equilibrium has not been reached in $200\,A\,\rm GeV$ S+S
collisions.

Since particle ratios in a grand canonical description are
parametrized by chemical potentials, we now introduce and investigate
the dependence on the chemical potentials $\mu_b$ for the baryon
number and $\mu_s$ for the strangeness. While $\mu_s$ can be
fixed by the condition of local strangeness neutrality ($\rho_s=0$,
corresponding to the initial state of the collision), $\mu_b$ at
freeze-out can be adjusted by fitting the experimental $p/\pi$ ratio.
We prefer $S/A$ instead of $\mu_b$ as a parameter, because it is an
invariant of the hydrodynamical evolution and thus can characterize
the EOS during the whole expansion.

\begin{table}[htb]
\begin{center}
\begin{tabular} {|c|c|c|c|c|c|} \hline
\multicolumn{2}{|c|}{\phantom{$\Big\vert_N$} experiment}
& \multicolumn{4}{c|}{global hydrodynamics} \\ \hline
			& NA35
	&                       & $S/A = 20$   & $S/A = 30$   & $S/A = 40$ \\
			\phantom{$\Big\vert_N$}
& total $y$
	&                       &$\!T_f\!\approx\!150\rm\,MeV$
                        &$\!T_f\!\approx\!162\rm\,MeV\!$
                        &$\!T_f\!\approx\!166\rm\,MeV\!$ \\ \hline\hline
\phantom{$\bigg\vert_N$}
$p / h^-$               & $0.28 \pm 0.04$
	& $p / \pi^-$           & 0.45          & 0.27          & 0.21 \\ \hline
\phantom{$\bigg\vert_N$}
$K^0_s / h^-$           & $0.10 \pm 0.02$
	& $K^0_s / \pi^-$       & 0.19          & 0.20          & 0.21 \\ \hline
\phantom{$\bigg\vert_N$}
$\Lambda / h^-$         & $0.080 \pm 0.010$
	& $\Lambda / \pi^-$     & 0.17          & 0.13          & 0.11 \\ \hline
\phantom{$\bigg\vert_N$}
$\antiLambda / h^-$     & $0.015 \pm 0.005$
	& $\antiLambda / \pi^-$ & $\approx0.005$&$\approx0.015$ &$\approx0.024$ \\
\hline
\end{tabular}
\begin{minipage}[t]{13.4truecm}
\caption[]{\sloppy \label{ratios}
	    The particle ratios from our model in comparison with the data
	    from NA35~\cite{NA35strange,NA35charged}. The actual number of
	    $\pi^-$ in the experiment could be smaller than the measured
	    negative hadrons $h^-$ by $\approx 5\%$ because of $K^-$ and $\bar
	    p$ contributions, whereas the protons are measured as the
	    difference between positive and negative hadrons.
	}
\end{minipage}
\end{center}
\end{table}

The particle ratios (including resonance decays) from global
hydrodynamics are shown in Tab.~\ref{ratios} in comparison with data
from NA35. We compare the total multiplicities, thereby averaging
over the rapidity distributions, which may -- especially in case of
the protons -- be different between our model and the experiment
(Fig.~5 in \cite{pheno}). Again the hydrodynamics results are almost
independent of the initial longitudinal fluid rapidity $\eta_0$. As
expected, the $p/\pi^-$ ratio varies strongly with $S/A$, and the
measured ratio can be reproduced quite nicely by the choice $S/A=30$.
If we would try to account for the reduced baryon content at central
rapidities, which is, as mentioned, not fully reproduced by our model
for the central zone, we would opt for a values of $S/A=40$.

The relative multiplicities of the strange particles are generally
overestimated by the model, which at first sight seems to indicate
that in the experiment chemical equilibrium for strangeness has not
been established yet by the $s\bar s$ generating processes. In all
three cases the value of $K^0_s/\pi^-$ is almost the same, because the
higher kaon abundance at higher temperatures is offset in the
denominator by the pion production from increased resonance decay
contributions. The $\antiLambda$ depend very strongly on the
freeze-out temperature and the baryon content.

However, the apparent overprediction of the strange particle ratios
relative to the negatively charged particles could also be due to an
underprediction of the total pion yield by the thermal equilibrium
model. As a number of authors have pointed out \cite{pionpotential},
the pion creation processes might dominate the reverse reaction for
some time leading to an overabundance of pions compared to their
chemical equilibrium value. Such non-equilibrium features can be
modelled by introducing a finite pion chemical potential $\mu_\pi$. It
would attain equal (positive) values for particles and antiparticles,
in contrast to the vanishing chemical equilibrium value for bosons
without conserved quantum numbers.

Tab.~\ref{pionmult} shows that global hydrodynamics lacks indeed both
in the number of pions and their energy. This lack of pion
multiplicity accounts for a large fraction of the strange/non-strange
discrepancy between theory and experiment seen in Tab.~\ref{ratios}.
Therefore the strange sector in S+S collisions is actually much closer
to chemical equilibrium than first suspected \cite{strangeequil}. This
was recently confirmed by a more thorough chemical analysis of the
data in \cite{chemfreezeout}.

\begin{table}[htb]
\begin{center}
\begin{tabular}{|lr|c|c|c|c|} \cline{3-6}
\multicolumn{2}{c|}{}           & experiment
	&\multicolumn{3}{c|}{global hydrodynamics} \\ \cline{3-6}
\multicolumn{2}{c|}{}           & NA35
	& $S/A = 20$            & $S/A = 30$            &$S/A = 40$     \\ \hline
$N_{\pi^-}$             &       & $93 \pm 5$
	&$55\ldots59$           &$62\ldots 67$          &$64\ldots71$   \\ \hline
$E_{\pi^-}$             &(GeV)  & $71 \pm 7$
	&$45$                   &$54$                   &$57$           \\ \hline
$E_{\pi^-}/N_{\pi^-}$   &(GeV)  & $0.76$
	&$0.76\ldots0.84$       &$0.81\ldots0.87$       &$0.80\ldots0.89$\\ \hline
 $S_{tot}/(3N_{\pi^-})$  &       &
	&$7.4$                  &$6.5$                  &$6.2$          \\ \hline
\end{tabular}
\begin{minipage}{13.4truecm}
\caption[] {\sloppy \label{pionmult}
	    The total multiplicity and energy of $\pi^-$ are not very well
	    represented by our model. However, the energy per pion is
	    described rather well. The total entropy per pion is considerably
	    higher than the value of $\approx 4$ for a pure pion gas,
    	reflecting the large abundance of heavy particles in the resonance
    gas. The data are from \cite{WenigD,Stroebele}.
	}
\end{minipage}
\end{center}
\end{table}

By computing the total energy per pion $E_\pi/N_\pi$, we deduce from
the relatively good agreement and the high value of $S_{\rm
tot}/N_\pi$ that the discrepancy is not so much a result of the shape
of the pion spectrum but merely of the general underrepresentation of
the pions in our model relative to strange particles. The observed
small deviation in the ratio $E_\pi/N_\pi$ could be easily due to the
pions in the lowest bin ($m_T-m_\pi\le 100\,{\rm MeV}/c^2$), which
strongly influence the multiplicity because of the exponential
character of the $m_T$-spectrum, and which we cannot reproduce in full
detail (Figs.~2\&7 in \cite{pheno}).

If we were to consider that all the missing pions are generated by
strong Bose condensation, we would find (using Boltzmann distributions
only) a pion chemical potential of $\mu_\pi=50\ldots 80\rm\,MeV$ as a
rough estimate, without taking into account the effect of a
correspondingly harder EOS on the hydrodynamics. This value of
$\mu_\pi$ is considerably smaller than $\mu_\pi=126\rm\,MeV$ in
\cite{pionpotential} and thus sufficiently far away from the Bose
divergence at  $m_\pi = 139\rm\,MeV$ so that it does not lead to a
substantial overpopulation at low momenta.

\section{Summary and Conclusions}
\label{summaryandconclusions}

We started from a previous analysis \cite{pheno} of hadronic
transverse mass and rapidity distributions from $200\,A\,\rm GeV$ S+S
collisions to estimate the maximum information content a hydrodynamic
analysis can possibly deliver. Using all available information on
both the initial conditions as well as the freeze-out conditions, we
still cannot determine the initial longitudinal flow, or equivalently,
the initial energy density. However, we find that regardless of the
initial flow all possible initial conditions finally lead to the same
temperature and transverse flow at freeze-out, which we model with a
dynamical freeze-out criterium. This is apparently a consequence of
the dynamical freeze-out in conjunction with the transverse expansion,
which in the late stages of hydrodynamics proves to be much more
important for the rarefaction than the longitudinal expansion.

A comparison with our phenomenological flow analysis shows then good
agreement with the data for a wide variety of initial conditions and
freeze-out criteria. However, while both hadron resonance gas and
plasma equation of state produce good agreement with the data, a
purely pionic equation of state generates too much transverse flow and
can thus be excluded for all hydrodynamic initial conditions.
Furthermore, from the fact that the data show a lower number of
strange particles than our model with chemical equilibrium would
produce, we deduce that chemical equilibration has not been reached.
On the other hand we determine that the chemical non-equilibrium of
pions is not as big as to allow for a $\mu_\pi$ close to the pion
mass, we estimate instead $\mu_\pi\approx 50\ldots80\,\rm MeV$.

\bigskip
\noindent
{\sectionfont Acknowledgments}

\medskip
\noindent
{\sloppy
E.~S.\ gratefully acknowledges support by the Deutsche
Forschungsgemeinschaft (DFG), the Alexander-von-Humboldt Stiftung and
the U.S. Department of Energy under contract numbers DE-AC02-76H00016
and DE-FG02-93ER40768. U.~H.\ gratefully acknowledges support by the
DFG, the Bundesministerium f{\"u}r Forschung und Technologie (BMFT),
and the Gesellschaft f{\"u}r Schwerionenforschung (GSI).  }



\begin{thebibliography}{99}


\def\etal{{et al.}}
\def\nuclphys{{\it Nucl. Phys. }}
\def\zphys{{\it Z. Phys. }}
\def\physrep{{\it Phys. Rep.\ }}
\def\physrev{{\it Phys.~Rev.\ }}
\def\physreva{{\it Phys.~Rev.~A}}
\def\physrevb{{\it Phys.~Rev.~B}}
\def\physrevc{{\it Phys.~Rev.~C}}
\def\physrevd{{\it Phys.~Rev.~D}}
\def\physrevlett{{\it Phys. Rev. Lett. }}
\def\physlett{{\it Phys. Lett. }}
\def\repprogphys{{\it Rep. Prog. Phys. }}
\def\annphysNY{{\it Ann. Phys. (NY) }}
\def\menton{{{\it ``Quark Matter 1990''}, J. P. Blaizot \etal (eds.),
        {\it Nucl. Phys.} {\bf A525} (1991) }}
\def\tucson{{{\it ``Hadronic Matter in Collision 1988"}, P. A. Carruthers
        and J. Rafelski (eds.), World Scientific, Singapore, 1989 }}
\def\quarkgluonplasma{{ {\it ``Quark--Gluon--Plasma''}, R. C. Hwa (ed.),
        Advanced Series on Directions in High Energy Physics, Vol. 6,
        World Scientific, Singapore, 1990}}
\def\borlaenge{{{\it``Quark Matter '93''}, Proceedings of the 10th
        International Conference on Ultrarelativistic Nucleus--Nucleus
        Collisions, E. Stenlund \etal (eds.), Borl{\"a}nge, Sweden,
        20.-24.~June 1993, to appear in {\it Nucl. Phys.} {\bf A} }}

\bibitem{Stoecker}
H. St{\"o}cker, W. Greiner, \physrep {\bf 137}, 277 (1986)

\bibitem{BEVALAC}
H. H. Gutbrod, A. M. Poskanzer, H. G. Ritter, \repprogphys {\bf 52},
1267 (1989) \\
H. A. Gustaffson \etal, \physrevlett {\bf 52}, 1590 (1984) \\
R. E. Renfordt \etal, \physrevlett {\bf 53}, 763 (1984)

\bibitem{Katajahydro}
H.~v.~Gersdorff, L. D. McLerran, M. Kataja, P. V. Ruuskanen,
\physrevd {\bf 34}, 794 (1986); \\
M. Kataja, P. V. Ruuskanen, L. D. McLerran, H.~v.~Gersdorff,
\physrevd {\bf 34}, 2755 (1986)

\bibitem{Katajaflow}
M.~Kataja, \zphys {\bf C 38}, 419 (1988)

\bibitem{Gersdorffhydro}
H.~v.~Gersdorff, \nuclphys {\bf A525}, 697c (1991)

\bibitem{Ornikhydro}
U. Ornik, F. Pottag, R. M. Weiner, \physrevlett {\bf 63}, 2641 (1989)

\bibitem{Katscherhydro}
R. Waldhauser, D. H. Rischke, U. Katscher, J. A. Maruhn, H. St{\"o}cker,
W. Greiner, \zphys {\bf C54}, 459 (1992).

\bibitem{Fermi}
E. Fermi, {\it Prog. Theor. Phys.} {\bf 5}, 570 (1950); \\
E. Fermi, \physrev {\bf 81}, 683 (1951)

\bibitem{Landau} L. D. Landau, {\it Izv. Akad. Nauk SSSR} {\bf 17}, 51
(1953); also published in {\it Collected Papers of L.D. Landau},
D. Ter Haar (ed.), Gordon and Breach, New York, 1965

\bibitem{Carruthershydro}
P. Carruthers, Minh Duong-Van, \physrevd {\bf 8}, 859 (1973)

\bibitem{Bjorken} J. D. Bjorken, \physrevd {\bf 27}, 140 (1983)

\bibitem{AGSdata}
G.~S.~F. Stephans, S. G. Steadman, W. L. Kehoe (eds.), {\it Heavy Ion
Physics at the AGS, HIPAGS '93}, Massachusetts Institute of Technology
report MITLNS-2158, Cambridge, Massachusetts, 1993

\bibitem{borlaenge}
\borlaenge

\bibitem{SPSdata}
NA35 Coll., A. Bamberger \etal, \physlett {\bf B184}, 271 (1987); \\
NA35 Coll., H. Str{\"o}bele \etal, \zphys {\bf C 38}, 89 (1988); \\
NA34 Coll., T. {\AA}kesson \etal, \zphys {\bf C 46}, 361 (1990); \\
WA80 Coll., R. Albrecht \etal, \zphys {\bf C 47}, 367 (1990)

\bibitem{pheno}
E.~Schnedermann, J.~Sollfrank and U.~Heinz, \physrevc {\bf 48}, 2462
(1994)

\bibitem{global}
E. Schnedermann and U. Heinz, \physrevc {\bf 47}, 1738 (1993)

\bibitem{NA35strange}
NA35 Collab., J. Bartke \etal, \zphys {\bf C48}, 191 (1990)

\bibitem{WenigD} S. Wenig, PhD. thesis, Universit{\"a}t Frankfurt
(1990), GSI-Report GSI-90-23

\bibitem{NA35charged}
NA35 Coll., J. B\"achler \etal, {\it Charged particle spectra in
central S+S collisions at 200 A GeV/c}, submitted to \physrevlett

\bibitem{Sinyukovflow}
Yu. M. Sinyukov, V. A. Averchenkov, B. L{\"o}rstadt, \zphys {\bf C49},
417 (1991)

\bibitem{resonances}
J. Sollfrank, P. Koch, U. Heinz, \physlett {\bf B252}, 256 (1990); \\
J. Sollfrank, P. Koch, U. Heinz, \zphys {\bf C52}, 593 (1991); \\
H. W. Barz, G. Bertsch, D. Kusnezov, and H. Schulz, \physlett {\bf B254},
332 (1991)\\
G. E. Brown, J. Stachel, and G. M. Welke, \physlett {\bf B253}, 315 (1991)

\bibitem{circumstantial}
E. Schnedermann and U. Heinz, \physrevlett {\bf 69}, 2908 (1992)

\bibitem{averageflow}
Please note that we use cylindrical symmetry and a quadratic
transverse velocity profile $\beta_r(r,t) = \beta_s(t) (r/R(t))^2$,
such that the average transverse flow velocity $\langle \beta_r
\rangle$ is related to the surface expansion velocity $\beta_s$ by
$\langle \beta_r \rangle = 0.5 \,\beta_s$. The numerical result thus
corresponds to average flow velocities $\langle \beta\rangle = 0.25-
0.3\,c$ at freeze-out.

\bibitem{Stroebele}
NA35 Collab., H. Str{\"o}bele \etal, in: \menton, 59c

\bibitem{HBT}
G. Goldhaber, S. Goldhaber, W. Lee, A. Pais, \physrev {\bf 120},
300 (1960); \\
A. Bamberger \etal (NA35 Kollaboration), \physlett {\bf B203}, 320 (1988)

\bibitem{UliMayer} U. Mayer, E. Schnedermann, U. Heinz,
\physlett {\bf B294}, 69 (1992)

\bibitem{nonequilib}
S.~Gavin, \nuclphys {\bf B351}, 561 (1991)

\bibitem{HydroLepton}
M. Kataja, J. Letessier, P. V. Ruuskanen, A. Tounsi,
\zphys {\bf C55}, 153 (1992)

\bibitem{BondorfGarpmanZimanyi}
J. Bondorf, S. Garpman, J. Zimanyi, \nuclphys {\bf A296}, 320 (1978)

\bibitem{Search}
K. S. Lee, U. Heinz, \zphys {\bf C43}, 425 (1989); \\
K. S. Lee, U. Heinz, E. Schnedermann, \zphys {\bf C48}, 525 (1990)

\bibitem{vanHove}
L. van Hove, \zphys {\bf C27}, 135 (1985)

\bibitem{UdoEOS} U. Ornik, F. W. Pottag, R. M. Weiner, in: \tucson,
p. 310

\bibitem{Statistical}
H.~Satz (ed.), {\it Statistical mechanics of quarks and hadrons},
Amsterdam, North-Holland, 1981.

\bibitem{chemequilib}
P. Koch, B. M{\"u}ller, J. Rafelski, {\it Phys. Rep.} {\bf 142},
167 (1986); \\
H. W. Barz, B.L. Friman, J. Knoll, H. Schulz, \nuclphys {\bf A484}, 661
(1988)

\bibitem{pionpotential}
J. L. Goity, M. Leutwyler, \physlett {\bf B228}, 517 (1989); \\
M. Kataja, P. V. Ruuskanen, \physlett {\bf B243}, 181 (1990); \\
S. Gavin, \nuclphys {\bf A544}, 459 (1992)

\bibitem{strangeequil}
N. J. Davidson, H. G. Miller, R. M. Quick, J. Cleymans, \physlett
{\bf B255}, 105 (1991)

\bibitem{chemfreezeout}
J. Sollfrank, M. Ga\'zdzicki, U. Heinz, J. Rafelski, {\it Chemical
freeze-out conditions in central S-S collisions at 200 A GeV}, \zphys
{\bf C61} (1994), in press


\end{thebibliography}
\end{document}